# Transverse Wind Velocity Recorded in Spiral-Shell Pattern

Hyosun Kim

Korea Astronomy and Space Science Institute, 776, Daedeokdae-ro, Yuseong-gu, Daejeon 34055, Republic of Korea
*Corresponding Author: H. Kim, hkim@kasi.re.kr

## Abstract

The propagation speed of a circumstellar pattern revealed in the plane of the sky is often assumed to represent the expansion speed of the wind matter ejected from a post-main-sequence star at the center. We point out that the often-adopted isotropic wind assumption and the binary hypothesis as the underlying origin for the circumstellar pattern in the shape of multilayered shells are, however, mutually incompatible. We revisit the hydrodynamic models for spiral-shell patterns induced by the orbital motion of a hypothesized binary, of which one star is losing mass at a high rate. The distributions of transverse wind velocities as a function of position angle in the plane of the sky are explored along viewing directions. The variation of the transverse wind velocity is as large as half the average wind velocity over the entire three dimensional domain in the simulated models investigated in this work. The directional dependence of the wind velocity is indicative of the overall morphology of the circumstellar material, implying that kinematic information is an important ingredient in modeling the snapshot monitoring (often in the optical and near-infrared) or the spectral imaging observations for molecular line emissions.

## 1. Introduction

The upgrade of the Hubble Space Telescope (HST) and the very recent commencement of the James Webb Space Telescope (JWST) science operations in the optical to infrared, as well as the stable operation of the Atacama Millimeter/submillimeter Large Array (ALMA) at longer wavelengths have enabled deeper searches for finer dusty and gaseous structures imprinted in the circumstellar media of evolved stars. These observations are now providing much more detailed information about their mass-loss history during their late evolutionary stage. Of particular interest is the recent frequent detection of shell patterns in the circumstellar envelopes of asymptotic giant branch (AGB) stars and their remnants in the outer parts of (pre)planetary nebulae ((p)PNe). As well described in e.g., Guerrero et al. (2020), the detection of multilayered shells of gas and dust has accumulated over the past few decades. Initially these structures were naively thought to arise as a result of episodic mass loss during the late evolutionary stage of AGB stars, but complex structures are now found which include spirals or fragmented arcs that are not concentric.

The patterns in the shapes of multilayered shells, rings, spirals, and arcs are investigated and used to derive the expansion velocity of the circumstellar wind matter. Balick et al. (2012) compared the HST images of a pPN CRL 2688 at $\sim 0.6\,\mu$m taken in 2002 and 2009, from which they concluded that a "translation" of the ensemble of all observed rings is a more favorable interpretation than one based on a "magnification" of the rings with time. This translation corresponds to a differential proper motion at a constant velocity of $\sim 20\,\mathrm{km\,s^{-1}}$ assuming a distance of 420 pc (Ueta et al. 2006). As an alternative, they derived a distance of 340 pc by assuming the transverse wind velocity that they measured ($\sim 0\farcs 07$ in 6.65 yr) to be $18\,\mathrm{km\,s^{-1}}$, which corresponds to the Doppler shift velocity measured from the $^{12}$CO absorption feature.

For the same pPN, Ueta et al. (2013) compared the HST images at 1995 and 2002 epochs to derive the position angle dependence of the transverse wind velocity of the apparent arcs. The derived velocities of arc segments were $\sim 10\,\mathrm{km\,s^{-1}}$ at, and near, the bipolar lobes and increased up to $\sim 30\,\mathrm{km\,s^{-1}}$ at the position angle of 30°–40° away from the bipolar lobes. The measured velocities showed negligible variation along the radius. The average value of these velocities is consistent with the single value derived by Balick et al. (2012).

Guerrero et al. (2020) performed a similar analysis to the arcs and fragmented ring-like features of the AGB star AFGL 3068 and two PNe NGC 6543 and NGC 7027. Specifically, they adopted the magnification method in contrast to the translation method of Balick et al. (2012). However, it was applied to the localized regions around the individual arcs, which may render the methodological issues as insignificant. Several rectangular regions across sharp arcs were selected per target, in which the count number standard deviations of the pixels were minimized when the transverse wind velocities of the individual arcs were properly computed. Although the number of samples for the velocity measurements per target was small, one indubitable finding was the very large velocity dependence on the position angle; the velocity variation was ≳14 km s$^{-1}$ in their measurements for these three sources (see Figure 5 in Guerrero et al. 2020).

Kim et al. (2021) chose a global analysis, as done by Balick et al. (2012), but the measurements were obtained independently for 36 sectors with a sector width of 10°. In their work, they ignored the radius dependence (as shown in the above papers) and adopted a statistical approach to the pattern segments within a sector. The resulting transverse wind velocities of the ring-like pattern of the AGB star IRC+10216, seen in the HST images in 2011 and 2016, vary from ≲10 km s$^{-1}$ (north) to ≳20 km s$^{-1}$ (south). Furthermore, the average of the transverse wind velocities was derived to be 12.5 km s$^{-1}$, which is about 2 km s$^{-1}$ lower than the wind speed measured along the line of sight based on using molecular line emissions. It was pointed out that the directional dependence of the wind expansion velocity can be understood in the framework of an eccentric binary model. Based on the generalized binary models in Kim et al. (2019), it was suggested that the projected location of the pericenter of the AGB star component in IRC+10216 may be toward the northern direction as indicated by the minimum transverse wind velocity (see also Kim et al. 2023).

The recent (sub)millimeter interferometric observations of molecular line emissions at high angular/spectral resolutions, in particular, have revealed multilayered shell patterns in the circumstellar media surrounding evolved stars (e.g., Maercker et al. 2012; Kim et al. 2017; Guélin et al. 2018; Decin et al. 2020). The desire for higher resolution studies is resulting in the discovery of multilayered shells that were previously unknown due to small intervals of the pattern segments. Since a small spacing of the pattern corresponds to a smaller orbital period in the binary-induced spiral-shell model (e.g., Soker 1994; Mastrodemos & Morris 1999; Kim & Taam 2012a) at a given range of terminal velocity of the wind, higher resolution observations will facilitate the identification of binary systems with lower mass (thus, possibly planetary) companions in the near future.

In this paper, we revisit the models for a comparable-mass binary that induce a spiral-shell pattern as the underlying theory for the anisotropic morphology and expansion velocity of the wind. An understanding and determination of the different fluid velocities toward different directions will provide important clues for the binary orbital elements such as the orbital shape, the inclination angle, the location of the line of nodes of the orbit, the location of the pericenter in a noncircular orbit, and the mass ratio.

As in a predominantly expanding system, the pattern speed that is measured from images taken at two different epochs corresponds to the average velocity of the expanding wind material that compose the pattern. For a slow wind where the terminal velocity of the wind expelled from a mass-losing star is comparable to the orbital velocity of the star, the wind material is directed around the binary stars and the pattern speed (averaged over the period between the two snapshot observations) may depart from the instantaneous expansion speed of the material. This paper focuses on sufficiently fast winds that are not dominated by the rotational motion of the binary system.

## 2. Method

We adopt the four models that are already developed by Kim et al. (2019). These models are based on the results of three dimensional (3D) hydrodynamic simulations using the code FLASH (Fryxell et al. 2000) in version 4.3 with adaptive mesh refinement. The calculations were performed in Cartesian coordinates with the origin at the center of mass of the binary system. The equation of state is that of an ideal gas with the ratio of specific heats, $\gamma$, chosen to be equal to 1.4. The system of hydrodynamic equations (continuity, momentum, and energy equations) was solved with the extra forces due to the gravity of the binary stars taken into account. In particular, the gravitational force attributed to the mass-losing star of mass $M_1$ located at $\vec{r} = \vec{r}_1$ with the gravitational softening radius of $\epsilon_1$ is given by

$$-\vec{\nabla}\Phi_1^{\rm eff} = -\hat{\mathbf{r}}\frac{GM_1}{|\vec{r}-\vec{r_1}|^2 + \epsilon_1^2} \times (1-f), \qquad (1)$$

where the acceleration factor $f$ represents the ratio of the outward force owing to radiation pressure on dust grains to the inward gravitational force. The gravitational force attributed to the companion star, denoted by subscript 2, is given by

$$-\vec{\nabla}\Phi_2 = -\hat{\mathbf{r}}\frac{GM_2}{|\vec{r}-\vec{r_2}|^2 + \epsilon_2^2} \times \mathcal{W}, \qquad (2)$$

where the parameter $\mathcal{W}$ is set to be 0 in their Models 1 and 3, and to be 1 in Models 2 and 4. Regardless of the value of $\mathcal{W}$, the orbit of the mass-losing star was preset according to the binary dynamics. Specifically, the same circular orbit in Models 1 and 2 and the same eccentric orbit with the eccentricity $e$ of 0.8 in Models 3 and 4 was adopted. In order to clearly examine the effects of companion's wake, a rather massive companion $M_2 = 2.2\,M_\odot$ was adopted. The current mass of the mass-losing star $M_1 = 0.8\,M_\odot$ was assumed constant, although the mass loss is expected to be nonnegligible (about 1%) during the simulation time (8 orbital periods, or ~2600 yr) for the adopted high mass loss rate of $\dot{M}_1 = 3.2 \times 10^{-6}\,M_\odot\,{\rm yr}^{-1}$. The average binary separation was 68 AU. For more details on the simulation setup, refer to Kim et al. (2019).

In order to consider all possible views of the simulated data cubes, we utilize the Euler angles defined by three elemental rotations, guaranteeing their adequacy to reach any desired target frame. The simulated data cubes that we present here correspond to the time that the stars return back to their initial locations (at the apocenter in eccentric orbit cases) after 8 orbits. The simulated data cubes are initially described in the coordinates with the coordinate center located at the center of mass of the binary system. These cubes are first horizontally shifted to place the mass-losing star at the coordinate center, which will be hereafter described as the original (fixed) coordinate system $XYZ$ with the orbital axis along $Z$ and the pericenter of the mass-losing star at $+X$ (see Figure 1). The rotating coordinate system $xyz$, overlapping the motionless $XYZ$ frame, first rotates about the $Z$ axis by the first Euler angle in the counterclockwise direction; it relocates the pericenter of the mass-losing star from $+x$ toward the $-y$ axis. The $xyz$ system, then, rotates about the $x$ axis by the second Euler angle, which we denote as the inclination angle with respect to the line of nodes of the binary orbits along the $x$ axis at this stage. If the first Euler angle is nonzero, the pericenter of the mass-losing star will drift in 3D space, in general, offset from the $xy$ plane (i.e., the plane of the sky as the observers are sitting on the $+z$ axis). The third Euler angle can be further applied to the $xyz$ system about the $z$ axis, which simply rotates the observed image; we do not explicitly apply the third Euler angle for figures shown in this paper.

For a given viewing coordinate system of the simulated data cube the density, temperature, and radial velocity[1] (about the mass-losing star) data is sliced in the midplane (i.e., $xy$ plane). An image obtained from observations may represent the integration of these quantities along the line of sight (e.g., column density), and the appropriate integration method could be determined through radiative transfer calculations. We confirm with some tests that the integration of some volume around the midplane does not change the major conclusion of this paper with regard to the trends in the directional dependence of the expansion velocity. For example, a linear integration of quantities over the entire simulation domain merely increases the dispersions in the expansion velocity measurements, compared to the usages of quantities in the midplane slices as in the adopted method.

The "pattern" in an observed image may be defined as the regions in threads having high contrasts in surface brightness over the inter-pattern regions, which can be ill-defined, in particular for the outer segments of the pattern, because of insufficient contrasts limited by the depth of observation. It is easier to define with the simulated data cubes of density, temperature, and velocity components, where the characteristic jump conditions for shock fronts at the inner and outer edges of the spiral-shell pattern (see e.g., Figure 6 of Kim et al. 2019) are preserved. We have noticed that the temperature, in particular, sharply rises at the inner and outer edges of the spiral-shell pattern, above tens of Kelvins in the four models, in contrast to the low inter-pattern temperature of the order of unity. We define the "pattern" of our Models 1–4 with the criteria[2] satisfying the temperature over 10 K and the distance from the mass-losing star over 1 kAU. The latter criterion is adopted to avoid over-weighting the central core region characterized by high density. We note that trials with different temperature criteria merely change the dispersions of the measured expansion velocities and do not change the major conclusions in this paper.

---
[1] The radial velocity in this paper denotes the transverse velocity component moving away from the mass-losing star. It is different from the velocity along the line of sight, which is also often referred to the same expression.
[2] The numeric values in the criteria for defining the pattern are limited to these particular models that we already know the physical quantities in the

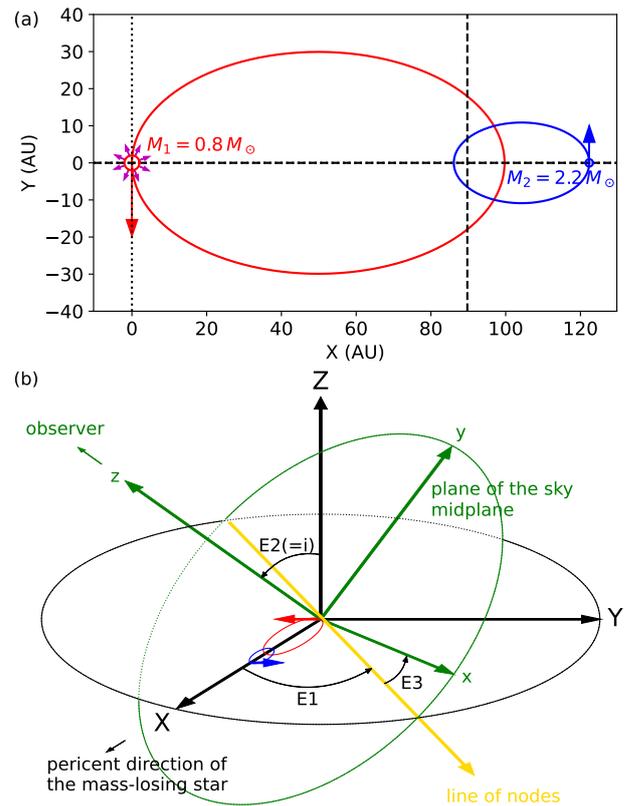

**Figure 1.** Schematic diagrams describing the coordinate systems and the locations of simulated binary orbits. (a) The mass-losing asymptotic giant branch star with the mass of $M_1$ and its companion with the mass of $M_2$ without mass loss move along red and blue ellipses, respectively. The simulations are performed in the coordinates with the origin at the center of mass of the binary (at the intersecting point of dashed lines). The $XY$ frame are shifted with its coordinate center to the current position of the mass-losing star, which is horizontally aligned with the center of binary mass at the present time, adopted for all figures displayed in this paper. The direction of motion of the star is indicated by an arrow with the same color for the position of each star and its orbital shape. (b) The coordinate transformation between the fixed $XYZ$ coordinates and the moving $xyz$ coordinates by rotations with three Euler angles denoted by E1, E2, and E3. The inclination angle, $i$, corresponds to E2.

Figure 2 displays the density-weighted average of the radial velocity components in expansion for 36 sectors with an angular width of $10°$. The standard deviation of the expansion velocities within the sector, separately calculated for the values above and below the average velocity, are denoted by the vertical bars. The colored curves in each panel show the inclination effects for each model. On the other hand, by comparing the profiles. In the analysis of observations, the brightness temperature does not increase as rapidly because it is an outcome of radiative transfer having a dependence on both density and (physical) temperature of the material.

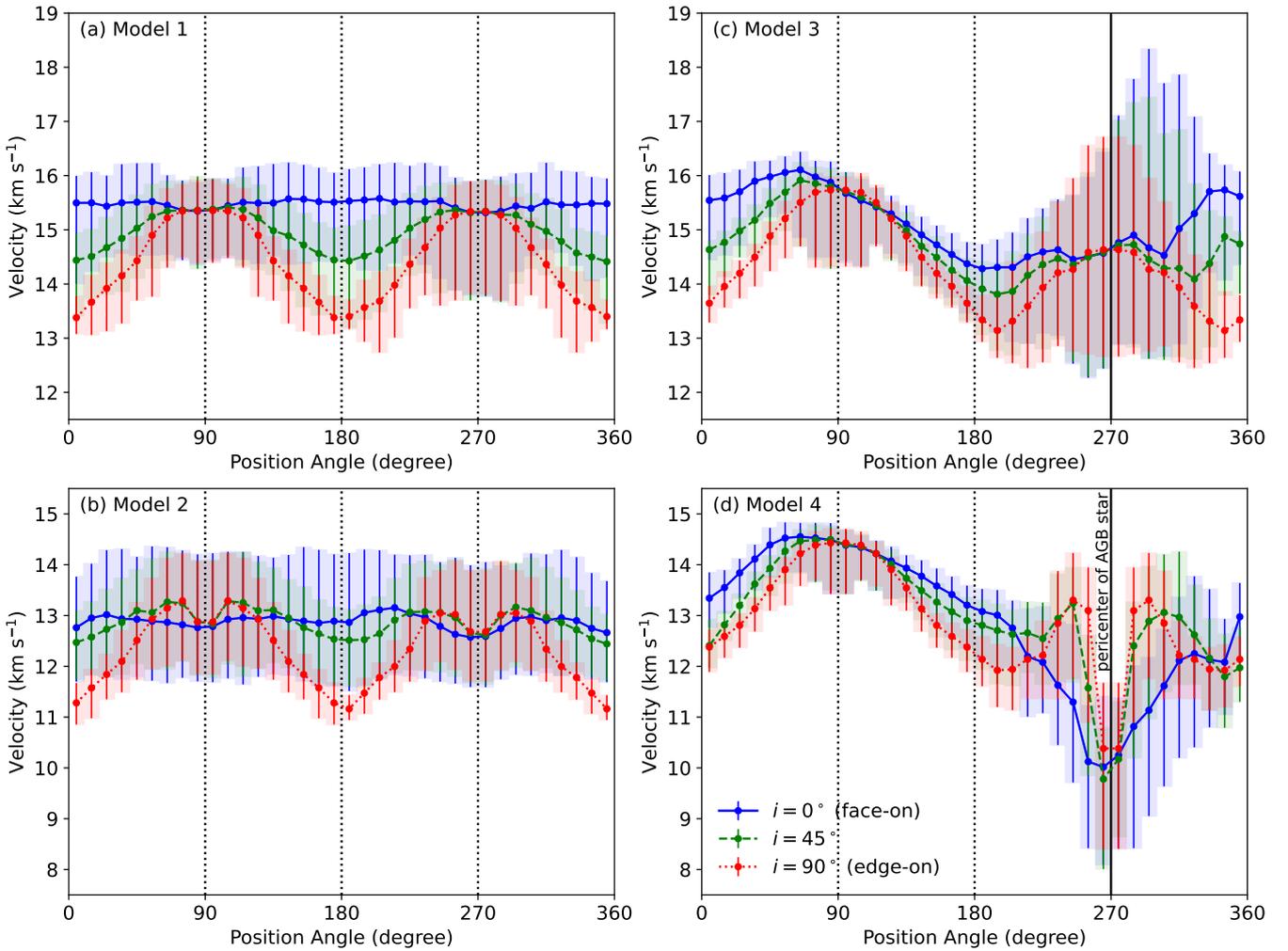

**Figure 2.** Measured expansion velocities of the spiral-shell patterns in (a) Model 1, (b) Model 2, (c) Model 3, and (d) Model 4, adopted from Kim et al. (2019), as a function of position angle that is measured from north ($+y$) to the east ($-x$). See Figures 3–6 for the coordinate information. Blue, green, and red symbols and lines indicate the measurements for cases of inclination angles of 0° (face-on), 45°, and 90° (edge-on), respectively. The pericenter of the mass-losing star is located to the west ($+x$ axis; position angle of 270°) in the orbital plane. The line of nodes of the binary orbits is $x$-axis, therefore the values in the curves for different inclination angles are the same at the position angles of 90° (east) and 270° (west), respectively. The vertical bar of each data point is the standard deviation of the velocities of the gaseous matter within the sector.

same colored curves across panels, the influence of eccentricity ((a) versus (c); (b) versus (d)) and of companion's wake ((a) versus (b); (c) versus (d)) are illustrated. The corresponding snapshots for temperature, density, and radial velocity are exhibited in Figures 3–6. The transverse wind velocity distributions in the cases that the pericenter is misaligned from the line of nodes of the orbital plane are presented in Figures 7 and 8.

## 3. Result

### 3.1. Views with the Pericenter along the Line of Nodes of the Orbit

In this section, the first and third Euler angles are set to be zero. Therefore, the pericenter of the mass-losing star is placed at $+x$ along the line of nodes of the orbit. The position angle is measured from the north ($+y$) to the east ($-x$), therefore the pericenter in this section is located at the position angle of 270°.

Model 1 is the ideal case to understand the basic geometry of the circumstellar medium induced by the orbital motion of a binary system. Figure 2(a) shows that the expansion velocity in the face-on view does not depend on the position angle, indicating the equal velocities in the orbital plane. For the views not exactly face-on, there are two peaks at the position angles of 90° and 270° along the line of nodes of the orbit, with valleys at the position angles of 0° and 180° being the deepest at the edge-on view. It shows that the speed in the orbital plane is faster than the speed along the orbital axis. We also note that the wind speed along the orbital axis coincides with the intrinsic wind velocity near the locus of the companion ($\sim 13.4$ km s$^{-1}$; see Kim et al. 2019). Some fraction of the speed of the stellar motion is added to the net velocity of the wind matter in the orbital plane. Therefore, the derived velocity distribution of the spiral-shell pattern as a function of the inclination and the wind speed ratio between the orbital axis and the orbital plane

**Transverse Wind Velocity**

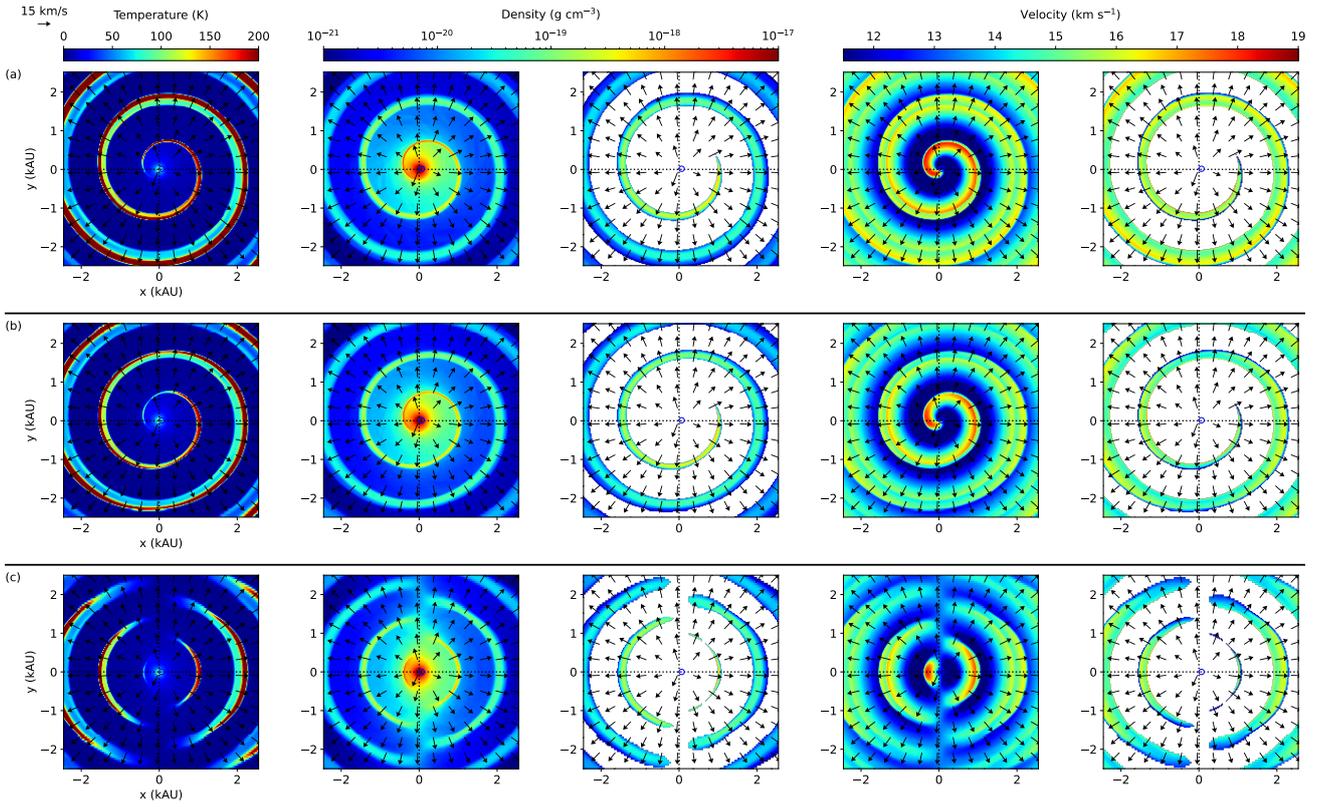

**Figure 3.** Temperature (first column), density (second and third columns), and radial expansion velocity (fourth and fifth columns) distributions of Model 1 at the midplane of the simulation data cube as viewed at the inclination angles of (a) 0° (face-on), (b) 45°, and (c) 90° (edge-on). The third and fifth columns show the quantities within the "pattern", satisfying the criteria for the temperature above 10 K and the distance from the mass-losing star greater than 1 kAU. The center of the images and the velocity vectors is at the location of the mass-losing star.

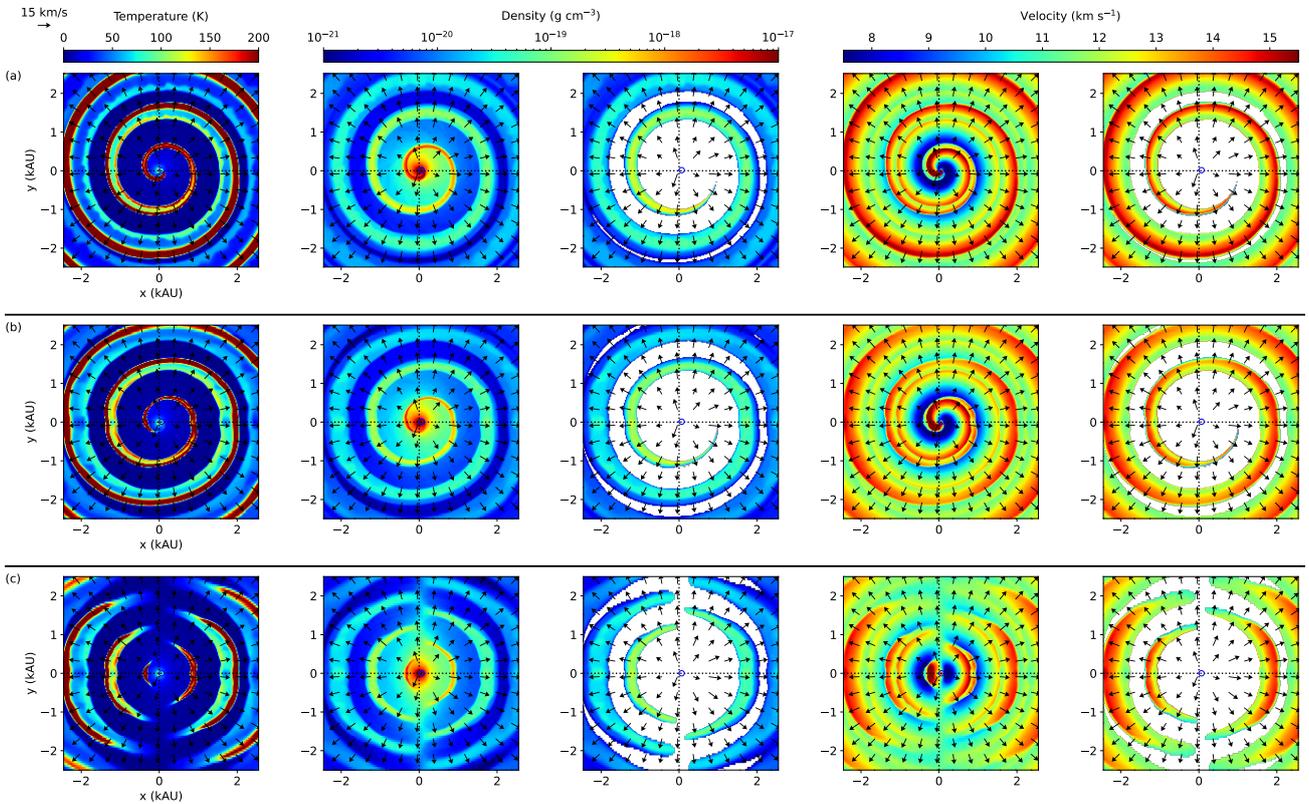

**Figure 4.** Same as Figure 3 but for Model 2.

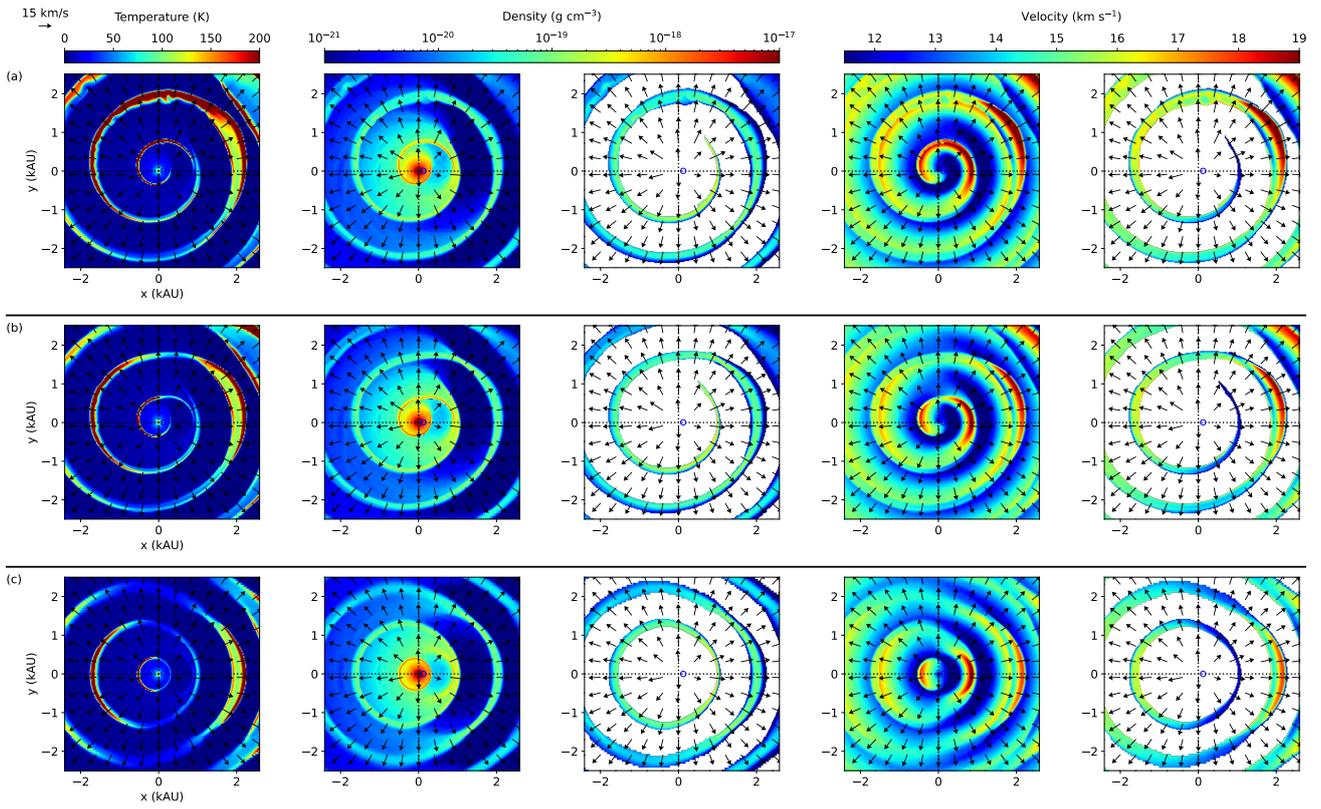

**Figure 5.** Same as Figure 3 but for Model 3.

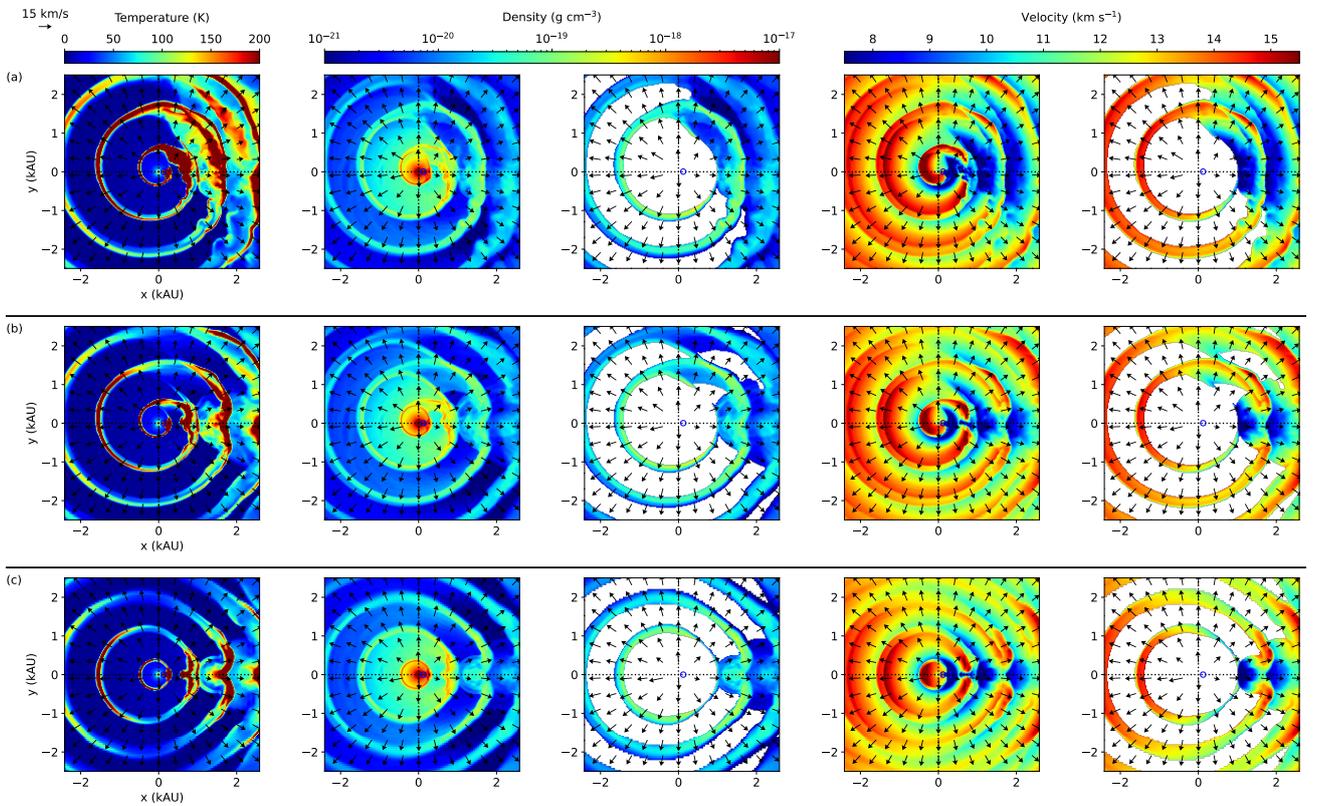

**Figure 6.** Same as Figure 3 but for Model 4.

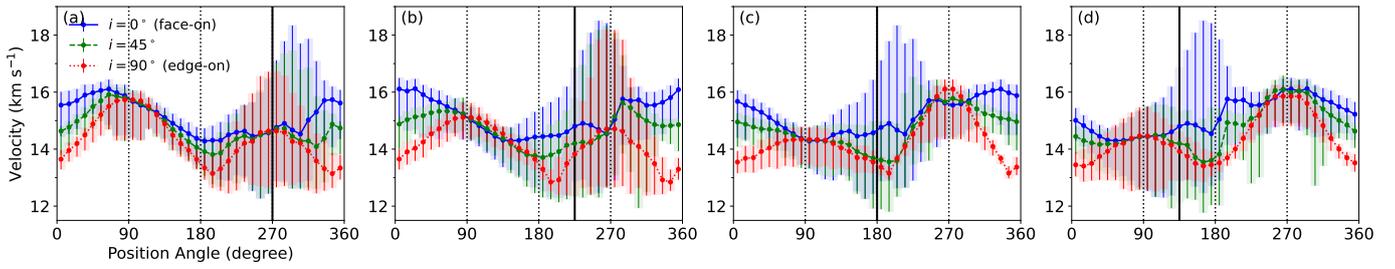

**Figure 7.** Effect of the first angle among the Euler angles for Model 3. The original data cube of simulation in *XYZ* coordinates with *Z* as the binary orbital axis is, first, rotated about the *Z*-axis by (a) 0°, (b) 45°, (c) 90°, and (d) 135°, relocating the pericenter of the mass-losing star (originally at +*X*) to the position angle (solid vertical line) of 270°, 225°, 180°, and 135°, respectively. And then, the rotated cube, shown in each panel with the above-mentioned first Euler angle being already applied, is further rotated by an inclination angle of 0° (blue), 45° (green), and 90° (red), respectively. The line of nodes of the orbital plane lies along the position angle of 90° and 270°. Total 12 combinations between the first and second Euler angles, covering all representative cases in the three dimensional angular space, are displayed. Introducing the third Euler angle will merely shift the three colored lines altogether along the position angle axis, which is not displayed here.

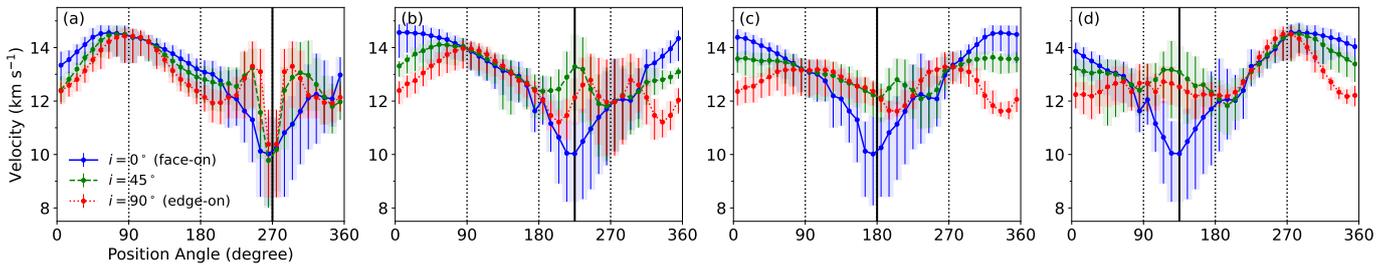

**Figure 8.** Same as Figure 7 but for Model 4.

(∼13.4/15.4) implies that the morphology of Model 1 is simply an oblate spheroid with about a 13% shorter axis aligned with the orbital axis. Figure 3(c), indeed, shows that the half circles of the pattern in the edge-on view are slightly elongated along the line of nodes of the orbit (i.e., *x*-axis) by about 13%.

Model 2 includes the gravitational potential of the companion star ($\mathcal{W} = 1$) and, thus, its gravitational density wake. The expansion velocities in this model are overall smaller by about 2 km s$^{-1}$ than in Model 1 in all 3D directions. Figure 2(b) interestingly illustrates that the velocities in the orbital plane (denoted in blue) are fluctuating. It is related to the companion's density wake overlaid on the main spiral-shell pattern. In the nonzero inclination cases, the clumpy higher-density structures appearing along the *x*-axis (i.e., the line of nodes of the binary orbit) indicate the cross sections of the companion's wake (see Figure 10 and the relevant text in Kim & Taam 2012a). The main spiral-shell pattern, similar to that in Model 1, has a vertical extent from the orbital plane to the angle described by the arc segments that are very close to the orbital axis (middle panel of Figure 4(c)). In contrast, the vertical extent of the companion's wake is very limited; the angular size of the arcs when viewed edge-on, that correspond to the cross sections of the companion's wake, is expected to be ∼10° only, based on Equation (13) of Kim & Taam (2012b). The overlap of these two spiral structures in the orbital plane causes the small variation in wind velocities, measured in the face-on snapshot. The off-orbital-plane velocities are also affected as shown in the green and red curves in Figure 2(b), in which the velocity peaks are slightly offset from the line of nodes of the orbit.

Model 3 and Model 4 are the cases for an eccentric orbit with and without the companion's wake, respectively. At the pericenter of the mass-losing star (position angle of 270°; +*x* direction), the orbital speed of the star and thus the alteration of the mass ejection velocity from the intrinsically isotropic value in the rest frame of the star are maximized. For this reason, the velocity variation near the pericenter becomes extreme (see Figure 2(c)–(d)). In addition, the greater wind expansion speed yields a stronger shock at the outer edge of the spiral pattern, enhancing the gas temperature, which enlarges the width of the pattern (see Figures 5–6).

In the face-on view for Model 3 the maximum velocity is not achieved at 270° but at about 300° (blue of Figure 2(c); see also the rightmost column of Figure 5(a)). This results from the fact that the matter is ejected at its highest speed toward the "forward" direction of the orbital motion when the mass-losing star was passing its pericenter at the position angle of 270°. At the same time, the minimum expansion velocity is detected also at the position angles around the pericenter of the mass-losing star, but near the inner edge of the spiral pattern where the gas pressure at the heated shock front acts in the opposite direction.

In Model 4, further complexities arise from the overlap of companion's wake in addition to the already intricate structure of an eccentric orbit model, in a manner similar to that described in Model 3. In particular, it causes turbulent eddies near the pericenter of the mass-losing star (see Figure 6(a)), resulting in the smallest average velocity in such a direction across the position angles (see Figure 2(d)).

## 3.2. Views with the Pericenter Located in 3D Space

In this section, we investigate the dependence of the image on the first Euler angle, which determines the position angle for the projection of the pericenter of the mass-losing star. Taken together with the inclination angle (second Euler angle), the elevation angle of the pericenter from the midplane is also determined. Here, the third Euler angle is fixed to be zero, keeping the line of nodes of the orbital plane along $x$-axis.

Figure 7 shows the velocity distribution in Model 3. Figure 7(a), with the first Euler angle of 0°, is the same as Figure 2(c). The other panels display the results measured in the cubes with the first Euler angle with 45° steps. The blue curve, with the second Euler angle of 0°, is simply shifted to the left by 45° each, from (a) to (d). In contrast, the green and red curves are entirely redistributed following the combinations of the first and second Euler angles. The green curve presents the case that the pericenter is misaligned with the datum planes. The position angles at which the velocity dispersion is the largest (i.e., the longest vertical bars in each panel) are close to, but not exactly on, the position angle for the projected pericenter location of the mass-losing star.

In the edge-on cases (red curves), the orbital plane is perpendicular to the plane of the sky and the pericenter is located on or above the line of nodes of the orbit, at the position angle of either 90° or 270°. In panel (a) of Figure 7, the pericenter is located along the $+x$ axis, and the largest velocity variation at the position angle of 270° is natural. For the case displayed in red color in panel (b), the pericenter lies above the $+x$ axis with the projected distance decreased by 30%. As a result, the velocity variation near the position angle of 270° becomes larger than the red-colored case in panel (a). The velocity distribution in the position angle between 0° and 180° is not significantly modified. In panel (c), where the pericenter is located along the line of sight passing through the coordinate center $(x, y) = (0, 0)$, the velocity variation is rather mild, equally in both eastern (0°–180°) and western (180°–360°) sides. The red curve in panel (d), where the pericenter is above the $-x$ axis at the same height as in panel (b), recovers a somewhat larger velocity variation, but in this case, in the eastern direction.

The velocity distribution of Model 4 differs from that of Model 3 having both maximum and minimum velocities across the orbital plane in the position angles near the direction toward the pericenter of the mass-losing star. Near the pericenter of Model 4, the matter velocity does not increase as much as that of Model 3 but, instead, the kinematic energy cascades to small turbulent eddies. Panel (a) of Figure 8 shows that the velocities in all inclinations significantly decrease at the position angle of 270°, where the pericenter is located. From panels (a) to (d), the face-on (blue) curve is simply shifted along the position angle axis, retaining one velocity valley at the position angle for the pericenter of the mass-losing star. In contrast, the other inclination cases (green and red curves) do not maintain the very deep velocity valley in the panels (b)–(d). This reflects the fact that the small velocity of the pattern is a quite localized feature around the direction toward the pericenter lying above the plane.

## 4. Conclusion

In this paper, four hydrodynamic models for the binary-induced spiral-shells surrounding a mass-losing star are reanalyzed to examine the directional dependence of their fluid velocities. Toward this end, we have examined the wind velocities in the entire 3D domain by measuring the transverse wind velocities in the midplane of the simulated data cube under the rotation of two directional angles.

It is found that the distribution of propagation speeds of a spiral-shell pattern is clearly representative of an oblate spheroidal morphology of a circumstellar envelope in the circular-orbit binary model. In this case, there are some fluctuations which are indicative of the relative strength of the companion's gravitational density wake. In contrast, in a highly eccentric-orbit binary model, the expanding velocity of the pattern results in a significant alteration of its distribution in a distinctive manner as reflected by an increase in the velocity dispersion near the position angle toward the projection of the pericenter of the mass-losing star. In addition to the broadened velocity dispersion, the velocity profile averaged over radius exhibits a characteristic one-peak profile along the position angle with the peak toward the apocenter of the mass-losing star in the eccentric orbit model, when the spiral is viewed face-on. A quantitative model comparison for the observed shape of the multilayered pattern as well as the position-angle dependence of its expansion velocity would provide probes for the orbital shape, the inclination angle, the location of the line of nodes of the orbit, the location of the pericenter, and the mass ratio.

The results in this paper can be immediately applied to the studies on differential proper motion of multiple rings surrounding AGB stars and also those remaining in pPNe and PNe for those objects in which their winds are sufficiently fast where there is little difference between the pattern speed and speed of the expanding matter. Previous studies based on optical images are often aimed at deriving the expansion velocity of the halo matter under the tacit assumption of an isotropic AGB wind. It is now clear that this assumption is inconsistent with the underlying binary hypothesis for the origin for multiple rings.

The strong position-angle dependence of the expansion velocity of the circumstellar patterns, revealed in AGB stars AFGL 3068 and IRC+10216 and in the halo components of (p)PNe CRL 2688, NGC 6543, and NGC 7027 (Ueta et al. 2013; Guerrero et al. 2020; Kim et al. 2021), can be qualitatively understood within the framework of binary models. The findings of this paper can also be used to facilitate the kinematic interpretation of spectral imaging data of molecular line emissions taken with millimeter/submillimeter interferometers. In particular, we note that most of the above sources have been observed with ALMA at high resolution, even at multiple epochs for some sources. A study using spectral imaging data at two different epochs with a proper time interval would quantitatively place constraints on their binary properties.

## Acknowledgments

We are grateful to the anonymous referee for fruitful comments and Ronald E. Taam for his detailed comments and suggestions that have improved the presentation and clarity of an earlier version of the manuscript. This research was supported by the National Research Foundation of Korea (NRF) grant (No. NRF-2021R1A2C1008928) and Korea Astronomy and Space Science Institute (KASI) grant (Project No. 2023-1-840-00), both funded by the Korea Government (MSIT).